# Ant-net: An Adaptive Routing Algorithm

Chandana M  
Amity University  
Sector 125, Delhi NCR, India

Sanjeev Thakur  
Amity University  
Sector 125, Delhi NCR, India

*Abstract: —*
*With the increasing demand and complexity of networks, factors such as balancing the load, improving the performance, reducing delay and finding the optimal path between nodes in a computer network have become crucial. The traditional routing approaches are not able to perform up to the mark as they don't take into account factors such as reducing delay and adaptive nature. Thus there is a need for more sophisticated techniques to meet the current network problems. The problem solving nature of ant colonies have provided a new approach which inspired to solve these network problems.*

*This paper focuses on Ant net algorithm based on Ant colony optimization which is an adaptive routing algorithm. In this approach, models of collective intelligence are transformed into optimization techniques. The ants travel across various paths in the network, thus depositing of pheromone, collecting route information and congestion. In this paper the Ant-Net algorithm along the important data structures required has been discussed and it is implemented on a simple packet switch network. The simulations describe the adaptive nature of the algorithm.*

*Index Terms: —* **Computer Networks, Adaptive routing, Ant colony optimization, Ant-net Algorithm.**

## I. INTRODUCTION

Day by day the importance of communication and internet is growing and more complex systems are being built. With the increase in complexity, there arises a need to address the various challenges that degrade the performance of the network [1].

A typical network consists of a number of routers and each router does communication with others. Each router can use some packets for navigation purpose and queues for buffering packets to each node. These packets should be distributed in such a way that the queue length and packet transfer time in each router can be reduced [3]. Some of the most common issues with routing in large networks are adaptability, load balancing, performance, congestion, selecting shortest and minimum cost path. The traffic in the network can be unpredictably high and also the structure of network changes according to the changes in requirement. So there aren't any constant parameters that can be used to route the network [11]. Thus there is a need for the development of new approaches that can route optimally thereby improving the efficiency of the system.

Computer Scientists researched and developed new routing algorithms using the behaviour of ants. Ants possess the ability of creating bi directional paths between their source and destination in minimum optimal amount of time as shown in figure 1. Ants deposit pheromone on their path and can sense pheromone deposited by other ants which helps them to guide towards the shortest path. And in case of obstacles, they are readily able to change their path. These properties of ants are what attracted the researchers towards them. By applying these abilities of the ants to routing algorithms, several network problems can be solved i.e. the internet flow can be made faster [12].

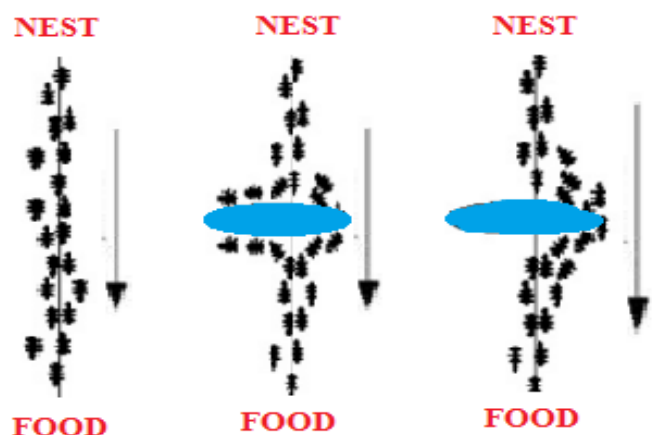

*Figure 1 Nature of ants*

Hence a new approach i.e. Ant Colony Optimization has been introduced by DiCaro and Dorigo in 1996 which is a metaheuristic that uses artificial ants to find desirable solutions for difficult optimization problems [9]. AntNet algorithm is an application of Ant Colony Optimization which has been discussed in the next sections.

## II. ANT NET ALGORITHM

The Ant Net algorithm has been proposed by DiCaro and Dorigo. It is a distributed and mobile agent's based algorithm which is used to solve routing problems. The artificial ants work together to obtain the solution in which stigmergy plays a very important role. Through indirect communication i.e. stigmergy, ants collect network information and using this information they build models of network status and pheromone tables [15].

In the Ant Net algorithm, each an every node in the network has the following data structures as shown in Figure 2:

**(i) Statistical parametric model:** It is a vector containing n-1 data structures where n is the total number of nodes. It is denoted by mean, variance and best travelling time over window of recent observations to reach the destination. The statistics model is used for the distribution of traffic to a specific destination [7].

**(ii) Pheromone Table:** The pheromone table is similar to the routing tables but instead of distances it consists of probability which helps us to choose the next node by telling the goodness of fit to reach a particular destination [4].

**(iii) Link Queues:** These queues are present at each node if the node has been provided with buffering capabilities irrespective of the algorithm. The priority of these queues is used by the backward and forward ant which is discussed in the algorithm [10].

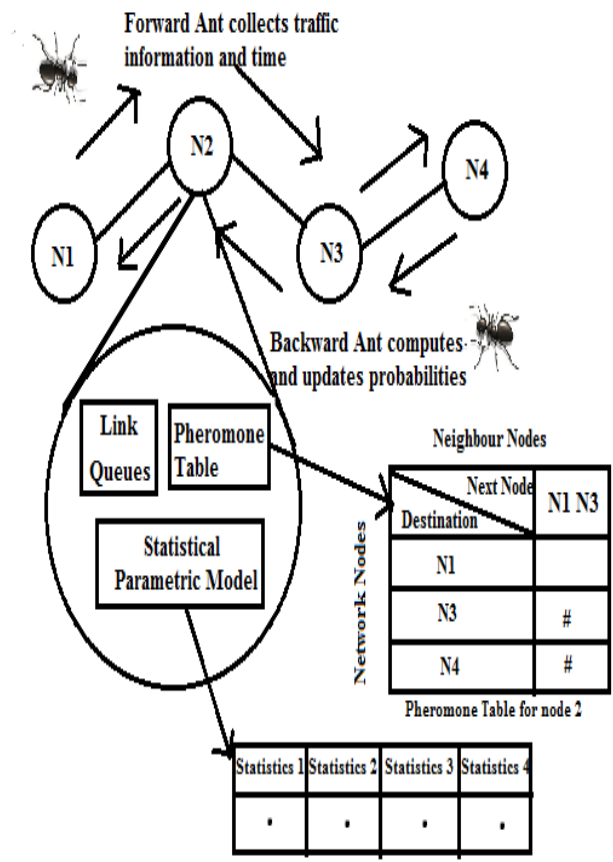

*Figure 2 Data Structures Used in Ant Net Algorithm at each node in a network [6]*

Now as we have discussed all the required data structure for the Ant Net, we move on to the algorithm. The working of the Ant Net Algorithm is based on the following two types of ant like agents:

☐ **Forward Ants:** The forward ants are used to grasp the information regarding the state of our network.

▪ **Backward Ants:** The backward ants are used to take the information and then update the routing tables of routers with that grabbed information which is already situated on network for further use [8].

The Ant Net algorithm can be described as the following:

1. In a network, each of the source nodes creates a forward ant and sends it to any random destination on regular time bases.

2. As the forward ant passes through each node, the time since the start of the ant is saved on a stack with an identifier that is used to identify the node as shown in the figure 3:

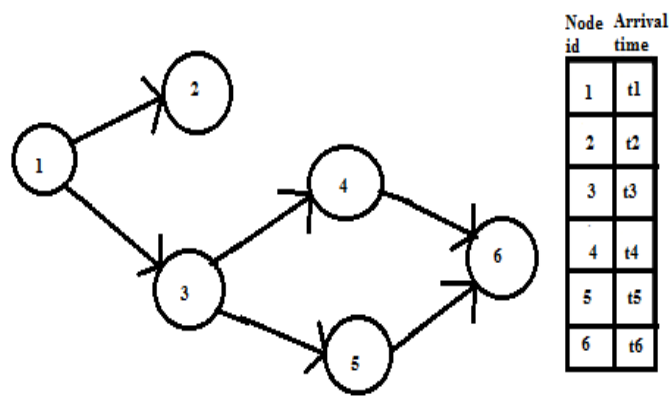

*Figure 3 Route travelling by forward ant [2]*

The Forward Ant has two ways of selecting the next hop which are as follows:

▪ It chooses among random nodes, where each random node has a probability to be selected.

▪ If the node it has selected in the previous way is already visited then it will choose another node having the same probability for all of its neighbouring nodes.

3. The node it has selected, if it was already visited then a cycle will be formed. Then in such a case the forward ant will pop all of the data from the stack. It's because an optimal path cannot have cycles. This is depicted in the following figure 4:

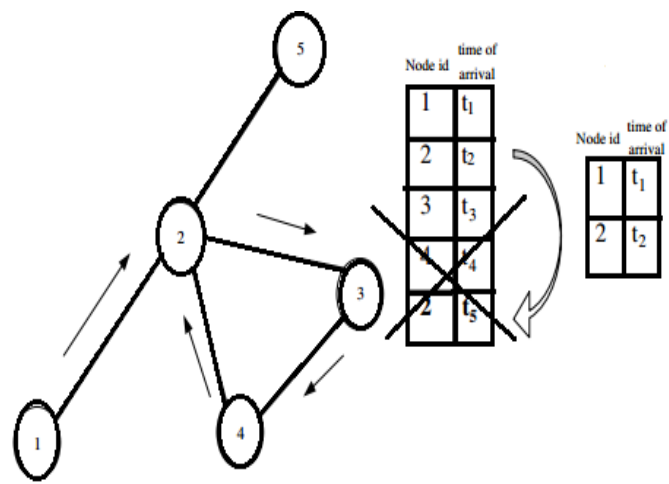

*Figure 3 Removal of loop by forward ant [10]*

4. While travelling, as the Forward Ant reaches the destination node then once again the time since the start of the ant along with the identifier of the node is stored on the stack of ant like agent. After this, now the Forward ant will be transformed into the Backward Ant and then transferring of the stack occurs. The Backward Ant thus returns to the source node in the opposite direction.

5. While reaching to the source node in the opposite direction, the Backward Ant will use the information that is found by the Forward Ants for obtaining updatations of the routing table through the path of the various nodes. As per the data contained in the stack, the probabilities associated to the path are used and this decreases the probabilities of other paths. The backward ant travels as follows for figure 3:

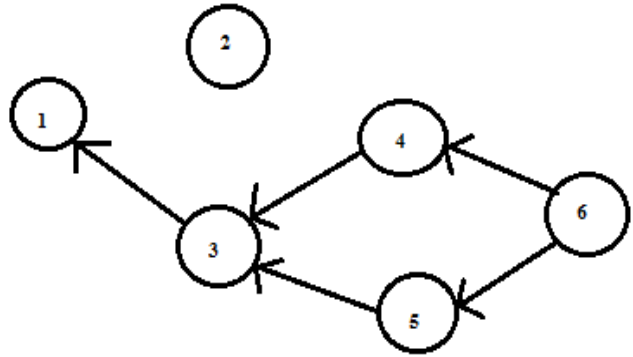

*Figure 4 Route travelling by backward ant [2]*

6. In the end when the backward ant reaches the source node and all of the updates are completed then it will stop working [13].

In this algorithm the priorities of the backward ant is more than that of the forward ants as the backward ants need to process as faster as it can so as to make the algorithm adaptive and due to this forward ants have to suffer from delays such that the algorithm has to avoid congestion. To balance the load, the time that has been saved onto the stack by the Forward Ants, are computed as the sum of two terms and not as the difference of the two time stamps where one of them represents the delay caused by the link load, and the other one represents the node load. The times when load is very high, the entity that represents the link node is thus increased for balancing the load [8].

III SIMULATION

The following topology of the network is used for simulation of the ant net algorithm. It has been projected in Microsoft Visual Studio 2010 [5].

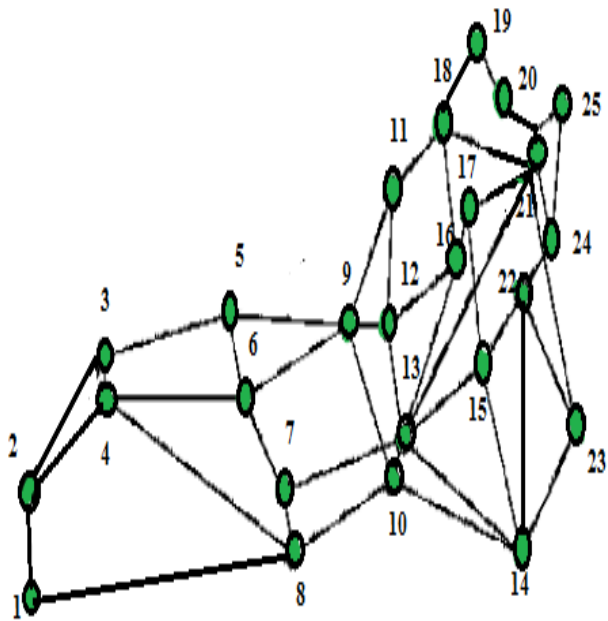

*Figure 5 Network Topology for Ant Net Algorithm Simulation*

The following shows the simulation result of the Ant Net algorithm after being performed on the above network:

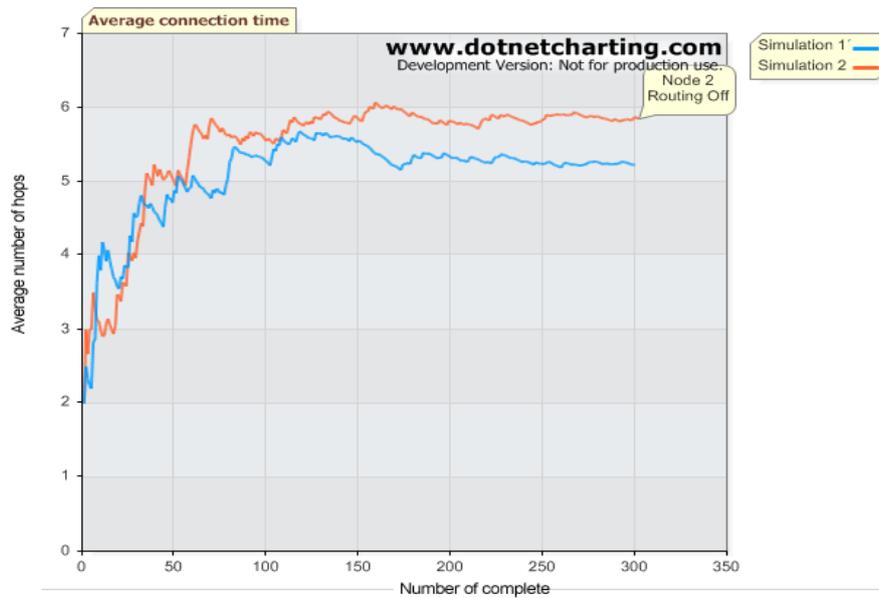

*Figure 6 Simulation of Ant Net Algorithm for 300 calls with removal of node 2*

In the above Figure 6,
**Simulation 1 Blue:** represents the normal run of the Ant Net algorithm without removal of any nodes.

**Simulation 2 Red:** represents the run of Ant Net algorithm with removal of node 2.

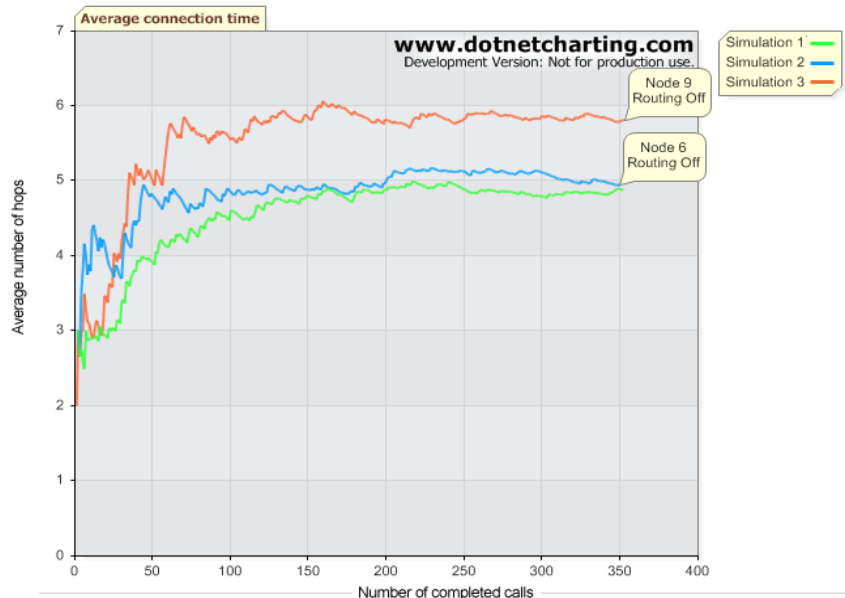

*Figure 7 Simulation of Ant Net Algorithm for 350 calls with removal of node 6 and node 9*

In the above Figure 7,
**Simulation 1 Green:** represents the normal run of the Ant Net algorithm without removal of any nodes.

**Simulation 2 Blue:** represents the run of Ant Net algorithm with removal of node 6.

**Simulation 3 Red:** represents the run of Ant Net algorithm with removal of node 6(removed previously in simulation 2) and node 9.

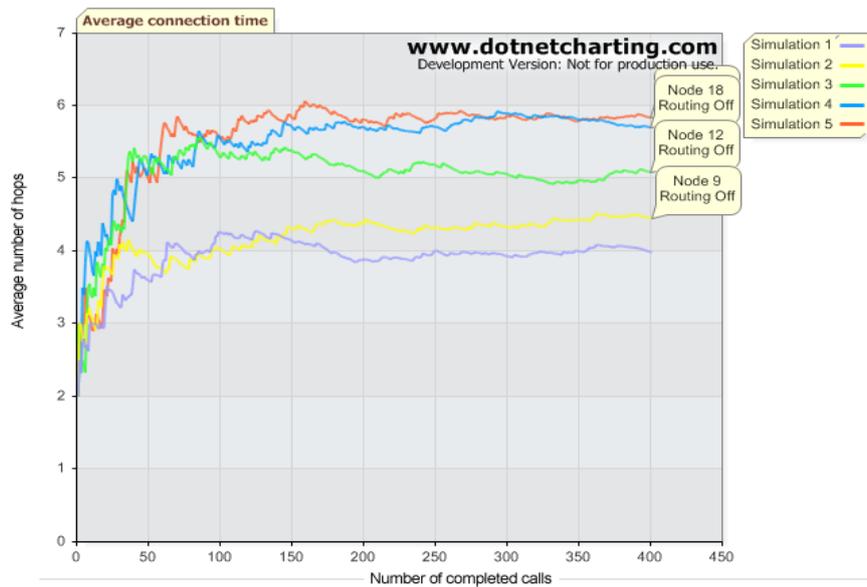

*Figure 8 Simulation of Ant Net Algorithm for 400 calls with removal of node 9, 12, 18 and 22*

In the above Figure 8,
**Simulation 1 Purple:** represents the normal run of the Ant Net algorithm without removal of any nodes.

**Simulation 2 Yellow:** represents the run of Ant Net algorithm with removal of node 9.

**Simulation 3 Green:** represents the run of Ant Net algorithm with removal of node 9(removed previously in simulation 2) and node 12.

**Simulation 4 Blue:** represents the run of Ant Net algorithm with removal of node 9, 12 (removed previously in simulation 3) and node 18.

**Simulation 5 Red:** represents the run of Ant Net algorithm with removal of node 9, 12, 18 (removed previously in simulation 4) and node 22.

## IV. CONCLUSION

In this paper we have discussed ant net algorithm, a relatively new approach for solving network problems. Simulation clearly shows that the Ant Net algorithm is adaptive in nature as even after removal of a number of nodes, it is able to complete all the calls successfully and thus helps in increasing the performance of the network. With the help of ant net algorithm we are able to find the optimal path, reduce the delay and improve the performance in the network.